# Squash root microbiome transplants and metagenomic inspection for *in situ* arid adaptations

Cristóbal Hernández-Álvarez[1,2], Felipe García-Oliva[3], Rocío Cruz-Ortega[4], Miguel F. Romero[1], Hugo R. Barajas[1], Daniel Piñero[5], and Luis D. Alcaraz[1*]

[1] Laboratorio de Genómica Ambiental, Departamento de Biología Celular, Facultad de Ciencias, Universidad Nacional Autónoma de México.

[2] Posgrado en Ciencias Biológicas, Universidad Nacional Autónoma de México.

[3] Instituto de Investigaciones en Ecosistemas y Sustentabilidad. Universidad Nacional Autónoma de México.

[4] Departamento de Ecología Funcional, Instituto de Ecología, Universidad Nacional Autónoma de México

[5] Departamento de Ecología Evolutiva, Instituto de Ecología, Universidad Nacional Autónoma de México.

*Correspondence author: lalcaraz@ciencias.unam.mx

# Graphical abstract

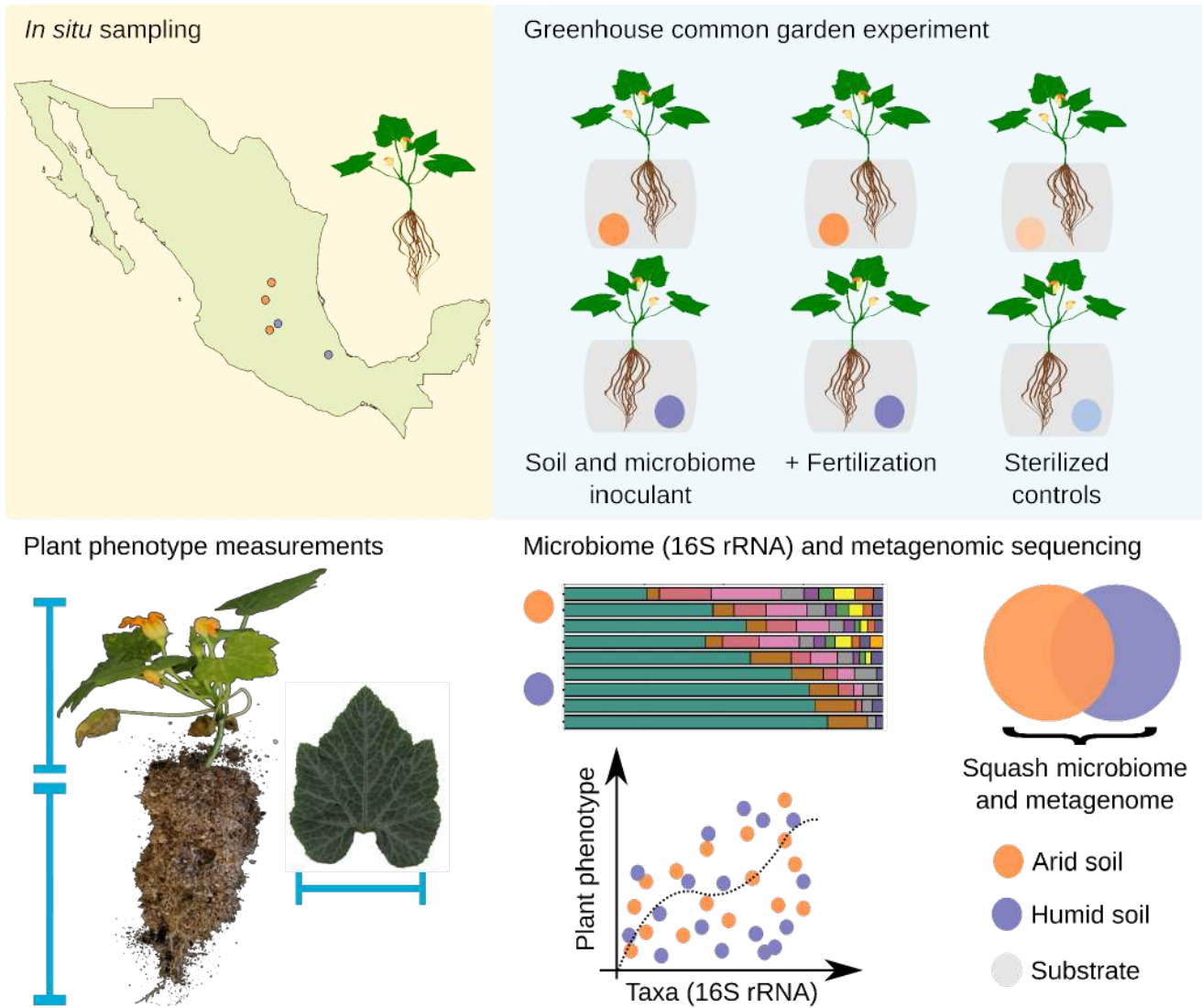

# Highlights

- We studied squash root microbiomes and their adaptations to arid environments.
- *In situ* collected soil was the primary microbial inoculant for a greenhouse common garden experiment.
- Bacterial diversity from the greenhouse resembled *in situ* bacterial communities.
- We described beneficial bacteria communities and their drought-resistance genes in squash roots.
- Microbes from arid soils are a promissory resource for plant management under drought.

## Abstract

Arid zones contain a diverse set of microbes capable of survival under dry conditions, some of which can form relationships with plants under drought stress conditions to improve plant health. We studied squash (*Cucurbita pepo* L.) root microbiome under historically arid and humid sites, both *in situ* and performing a common garden experiment. Plants were grown in soils from sites with different drought levels, using *in situ* collected soils as the microbial source. We described and analyzed bacterial diversity by 16S rRNA gene sequencing (N = 48) from the soil, rhizosphere, and endosphere. Proteobacteria were the most abundant phylum present in humid and arid samples, while Actinobacteriota abundance was higher in arid ones. The -diversity analyses showed split microbiomes between arid and humid microbiomes, and aridity and soil pH levels could explain it. These differences between humid and arid microbiomes were maintained in the common garden experiment, showing that it is possible to transplant *in situ* diversity to the greenhouse. We detected a total of 1009 bacterial genera; 199 exclusively associated with roots under arid conditions. By 16S and shotgun metagenomics, we identified dry-associated taxa such as *Cellvibrio, Ensifer adhaerens,* and *Streptomyces flavovariabilis.* With shotgun metagenomic sequencing of rhizospheres (N = 6), we identified 2,969 protein families in the squash core metagenome and found an increased number of exclusively protein families from arid (924) than humid samples (158). We found arid conditions enriched genes involved in protein degradation and folding, oxidative stress, compatible solute synthesis, and ion pumps associated with osmotic regulation. Plant phenotyping allowed us to correlate bacterial communities with plant growth. Our study revealed that it is possible to evaluate microbiome diversity *ex-situ* and identify critical species and genes involved in plant-microbe interactions in historically arid locations.

# 1. Introduction

One of the most significant challenges for the agricultural industry is climatic change, which can significantly affect yields in tropical areas as temperatures increase (Porter et al., 2014; Rosenzweig et al., 2014). Besides, climatic change may shrink areas with temperate and humid climates by increasing arid zones and minimizing the extension of lands suitable for agriculture at lower latitude locations (Dai 2013; Zarch et al., 2017; Zhang and Cai 2011). In this regard, to face problems that will result from climate change, organisms adapted to adverse conditions have the potential to be useful (Alsharif et al., 2020; Porter et al., 2014).

The bacterial diversity of soils is driven by biotic and abiotic factors that lead to spatial heterogeneity in the distribution of soil organisms (Comeau et al., 2020; Ettema and Wardle, 2002). Some abiotic parameters affecting the structure of the soil bacterial communities are pH, nutrient availability, salinity, soil texture, and aridity (Kaiser et al., 2016; Neilson et al., 2017; Peralta et al., 2013). Desert soils are ecosystems harboring a diverse set of bacterial species capable of surviving under drought conditions (Heulin et al., 2017; Lebre et al., 2017; Naylor and Coleman-Derr, 2018). As arid soils are oligotrophic ecosystems, soil microbial communities have a greater abundance of genes involved in the degradation of recalcitrant carbon sources (Naylor et al., 2017). Some soil bacteria can interact with plant roots, specifically within the rhizosphere and root endosphere, and the rhizosphere is the soil region directly in contact with plant root secretions (Bulgarelli et al., 2013; Sasse et al., 2018). The root endosphere is the microbial niche inside plant roots (Bulgarelli et al., 2013; Sasse et al., 2018). Given the observed pattern of reduction in bacterial diversity from soil to root (Comeau et al., 2020; Lundberg et al., 2012), a two-step microbiome selection model emerged (Bulgarelli et al., 2013). In the first step, soil properties shape microbial communities; in the second step, the plant selects its root-interacting bacteria by molecular signs as plant root exudates and the bacterial ability to interact with the plant cell wall host immune system (Bulgarelli et al., 2013). Several studies have described bacterial diversity from roots of plants in drought conditions, showing an increase in the abundance of Actinobacteriota (Fitzpatrick et al., 2018; Naylor et al., 2017; Naylor and Coleman-Derr, 2018; Santos-Medellín et al., 2017; Xu et al., 2018) which could be actively be recruited by root exudates (Xu et al., 2018).

Although not all interactions between bacteria and roots are beneficial for plants (Müller et al., 2016); some microbes from arid lands have the potential to facilitate plant growth under harsh conditions by improving plant health and enhancing resistance to high temperatures, low humidity, and elevated concentrations of salts (Alsharif et al., 2020; Soussi et al., 2016). Physiological adaptations to drought lead to differences between bacterial communities from deserts and other terrestrial ecosystems (Fierer et al., 2012; Neilson et al., 2017). Generally speaking, Acidobacteriota, Proteobacteria, and Planctomycetota bacterial phyla are associated with wet soils, while Actinobacteriota increases with aridity (Neilson et al., 2017). Actinobacteriota remains highly abundant in soils from arid localities because they are Gram-positive microbes with thick cell walls that tolerate desiccation (Naylor et al., 2017; Uroz et al., 2013; Xu et al., 2018).

In this work, we analyzed the root-associated microbiome (16S rRNA gene) and selected shotgun metagenomes of squash (*Cucurbita pepo* L.) in sites with different aridity levels to obtain microbial communities with the potential to improve plant survival to drought conditions. We studied squash plants because of their agricultural, economic, and gastronomic relevance (Eguiarte et al., 2018), but our experimental design can be applied to other plant species. We aimed to study and recover local diversity from arid ecosystems via a common garden experiment. The common garden experiment is a medium to study local adaptation, where individuals from different populations are grown in a common environment to study the relationships between phenotype, genotype, and environment (de Villemereuil et al. 2016). The existence of contrasting environments favors the study of local adaptation, and a common garden is a tool that allows researching the genetic basis of the phenotypes regardless if they are adaptive or not (de Villemereuil et al., 2016). We consider the microbiome as an extended plant phenotypic plasticity, with implications to plant local adaptation to its environment (Vannier et al., 2015; Partida-Martínez and Heil 2011; Y. Li et al. 2018). Here, we collected microbial communities from different habitats and then transplanted them to the same environment, a greenhouse, a common garden in a microbiological context (Reed and Martiny et al., 2007). Our common garden experiment evaluated the relationship between the microbiome and the plant's phenotype by growing squash plants in pots inoculated with the transplanted microbial communities. Finally, shotgun metagenomic analyses suggest that root-associated communities from arid sites possess genes to establish plants' interactions and survive drought.

## 2. Materials and Methods

2.1 *In situ* sampling

We chose five localities with active *Cucurbita pepo* farming (CONABIO, 2018), belonging to three soil groups, and due to their contrasting humidity climatic information (Table 1) (Fernández Eguiarte et al., 2014). We calculated Lang's aridity index (AI) (Quan et al., 2013). AI is the ratio of annual precipitation to mean annual temperature. Following their AI values, we classified the localities as arid (AI < 40) or humid (AI > 40) (Quan et al., 2013) (Table 1). We collected from the *in situ* cultivated, *C. pepo* plants roots (triplicates), and bulk soil samples collected to fill 50 mL sterile centrifuge tubes from these localities. Samples were immediately liquid nitrogen frozen and stored at -80ºC in the laboratory until metagenomic DNA extraction. Additionally, we collected 1 kg of soil at 10 cm depth for a common garden experiment, and we kept the soil at 4ºC until greenhouse use.

**Table 1.** Sampling locations and physicochemical characterization of soils.

|  | *In situ* location | AI | Latitude | Longitude | Soil order | pH | TOC (mg/g) | TN (mg/g) | TP (mg/g) | $NH_4$ (μg/g) | $NO_3$ (μg/g) | $HPO_4$ (μg/g) | Collection date |
|---|---|---|---|---|---|---|---|---|---|---|---|---|---|
| Humid localities | 1 | 99.4 | 18.81 | -97.08 | Vertisol | 7.1 | 26.41 | 2.12 | 0.75 | 0.35 | 46.17 | 100.36 | 2016/08/08 |
|  | 2 | 50.8 | 20.39 | -100.28 | Feozem | 6.9 | 18.61 | 1.39 | 0.316 | 0.36 | 60.55 | 148.72 | 2016/06/30 |
| Dry localities | 3 | 38.3 | 20.06 | -100.82 | Vertisol | 7.7 | 44.98 | 4.62 | 3.019 | 0.01 | 150.19 | 975.18 | 2016/06/30 |
|  | 4 | 24.7 | 21.59 | -101.08 | Feozem | 8.5 | 50.72 | 3.52 | 0.998 | 0.36 | 60.55 | 148.72 | 2016/06/29 |
|  | 5 | 21 | 22.45 | -100.70 | Xerosol | 8.5 | 12.39 | 0.87 | 0.082 | 0.34 | 88.58 | 317.68 | 2016/06/28 |

Sampling locations are sorted according to the aridity index (AI). Total organic carbon, TOC; total nitrogen, TN; total phosphorus, TP.

## 2.2 Soil processing and physicochemical description

Soil samples were cleaned for large pieces of organic matter and dried in an oven at 75°C. Then, we mixed the soil with deionized water 1:2 (w/v), and the pH was measured. Total and inorganic C were determined via a colorimetric method (Huffman, 1977) using a total carbon analyzer (UIC Mod. CM5012). Total organic carbon (TOC) was determined by subtracting levels of inorganic C from TOC. Total nitrogen (TN) and phosphorus (TP) were quantified using a Bran-Luebbe AutoAnalyzer III (Norderstedt, Germany) following previously reported protocols (Bremner and Mulvaney, 1982; Murphy and Riley, 1962). Inorganic fractions of N ($NH_4^+$ and $NO_3^-$) and P ($H2PO_4^-$) were estimated using standard methods (Murphy and Riley, 1962).

## 2.3 Common garden experiment and plants management

The common garden experiment considered reducing plant host genetic variance using commercial *Cucurbita pepo* var. Zucchini F1 seeds (Hydro Environment, http://hydroenv.com.mx) minimizing environmental variance by growing them simultaneously in a greenhouse under the same conditions. We tested *in situ* collected soils (N = 5) from arid and humid conditions as microbial inoculants for plant rhizospheres. For each tested soil, germinated plants were grown in a pot filled with 90% autoclaved sterilized substrate (agrolite and vermiculite 1:1 v/v) and a 10% v/v soil inoculum (N = 8 pots), the samples were distributed as follows: testing soil inoculants (n = 3 pots), fertilized soil inoculants (n = 3 pots), sterilized soil controls (n = 2 pots), and external control without inoculated soil (n = 1 pot).

Plant seeds were previously treated with thiram fungicide by the provider. We removed the fungicide and sterilized seeds by washing with sterilized water, 70% ethanol (v/v) for 5 min, 4.5% sodium hypochlorite (w/v) for 10 min, and performed a final wash in deionized water. Seeds were placed on Petri dishes using humid cotton and were maintained inside a germination chamber in dark conditions at 25.5°C until plant cotyledons were visible. The plants were watered daily. The fertilized treatment was initiated 34 days post-germination to test the role of external fertilization in the plants and microbiomes. According to the manufacturer (Agrodelta, http://www.agrodelta.com.mx/), the fertilization treatment consisted of weekly doses of commercial fertilizer containing 18.2 mg $NH_4^+$, 20.5 mg $P_2O_5$, and 35.1 mg $K_2O$. Sterilization controls were obtained by autoclaving the inoculant soils. The common garden experiment ended 52 days after transplantation (Fig. S1).

## 2.4 Plant phenotype measurements

We measured foliar surface, aerial plant length, stem length, flower number, root length, stem diameter, chlorophyll, carotenoids, aerial biomass, and specific leaf area. The absorbance of each solution at 663, 645, and 440.5 nm was measured using a spectrophotometer (Cary 50, Varian) to estimate chlorophyll a, b (Sudhakar et al., 2016), reported as chlorophyll, and carotenoid concentrations, as described elsewhere (Lewandowska and Jarvis, 1977). The foliar surface of specimens was determined by scanning leaves and analyzing pixel areas with ImageJ 1.x software (Schneider et al., 2012). Plant phenotype variables, including total chlorophyll and carotenoids, were measured by multiplying each pigment concentration by the plant's foliar surface area. Aerial biomass was estimated with plants dried in an oven at 50°C. Specific leaf area was calculated as the ratio of the leaf surface area to its dry weight (Wellstein et al., 2017).

## 2.5 Metagenomic DNA extraction

The metagenomic DNA was processed as previously described (Lundberg et al., 2012; Barajas et al., 2020; Romero et al., 2021). Briefly, plant roots were fractionated and collected into tubes that contained 1.5 mL of phosphate-buffered saline solution (PBS) at 7.5 pH. Tubes were vigorously mixed with a vortex until soil particles detached from roots to produce the root rhizosphere. For endosphere metagenomic DNA extraction, roots were mechanically broken and transferred to fresh tubes containing PBS (1.5 mL) and sonicated. Samples were centrifuged at 1,300 x g for 10 min to recover the rhizosphere and endosphere. Then, the metagenomic DNA was extracted using 0.25 g from the soil, rhizosphere, and endosphere pellets using the *MoBio PowerSoil® DNA Isolation Kit* (MoBio Laboratories, Solana Beach, CA, USA) following the manufacturer's protocol.

## 2.6 16S rRNA gene amplicons and metagenome sequencing

We performed PCR amplification of the 16S rRNA gene using MiSeq341F and MiSeq805R primers, corresponding to V3–V4 regions (Herlemann et al., 2011). The amplicon also included overhangs for multiplex library preparation by the Illumina Miseq protocol (Illumina Inc., 2013). Duplicate reactions were carried out for each sample and Taq polymerase (Promega. Madison, WI, USA). The amplification program involved an initial denaturation step at 94°C for 3 min; followed by five cycles that included denaturation at 94°C for 10s, annealing at 55°C for 20 s, extension at 72°C for 30 s; and 25 cycles of denaturation at 95°C for 5 s and extension at 72°C for 30 s. Amplicons were purified using the Wizard® SV Gel and PCR Clean-up System (Promega. Madison, WI, USA).

Sequencing of the amplified region of the 16S rRNA gene was performed using Illumina® MiSeq™ (Illumina, San Diego, CA, USA) hardware using a 2 x 300 bp configuration, following the Illumina® protocol provided by the *Unidad de Secuenciación Masiva* at the National Autonomous University of Mexico's Biotechnology Institute (IBT, UNAM). The whole shotgun metagenomic sequencing was performed at the *Laboratorio Nacional de Genómica para la Biodiversidad* (UGA-Langebio), using Illumina® NextSeq 500 (Illumina, San Diego, CA, USA) in a 2 x 150 bp configuration.

## 2.7 Sequence processing

The 16S rRNA gene analysis protocol used for this study had been previously reported and detailed at GitHub (https://github.com/genomica-fciencias-unam/SOP; Alcaraz et al. 2018; Hernández et al. 2020). Briefly, raw amplicon sequences were quality checked and trimmed to a length of 250 bp using FASTX-Toolkit (Gordon and Hannon, 2010). Sequences were merged with Pandaseq (Masella et al., 2012). Chimeras were removed using ChimeraSlayer (Caporaso et al., 2010). Sequence clustering of operational taxonomic units (OTUs) was performed with cd-hit-est (Li and Godzik, 2006) using an identity cut-off of 97%. Representative OTUs and taxonomically assignment were achieved with QIIME scripts (Caporaso et al., 2010) and BLAST (Altschul et al., 1990) the taxonomic database was SILVA v138 (Quast et al., 2013).

Raw metagenomic sequences were mapped against the *C. pepo* genome (Montero-Pau et al., 2018) using Bowtie 2 (Langmead and Salzberg, 2012) and were subsequently processed using SAMtools (Li et al., 2009) to filter out plant reads. Read quality control was done using Trimmomatic (4 nt sliding windows, Phred ≥ 15, 34 bp length) (Bolger et al., 2014). Metagenomic assemblies were done with metaSpades (Nurk et al., 2017) and Velvet (Zerbino, 2010) with the k-mer size equals to 31. Coding gene sequences and predicted proteins were calculated with Prodigal (Hyatt et al., 2010). Sequences of predicted proteins were aligned with DIAMOND (Buchfink et al., 2014) against the M5nr database (Wilke et al., 2012), using RefSeq (O'Leary et al., 2016), Subsystems ontology from SEED (Overbeek et al., 2005), and the Kyoto Encyclopedia of Genes and Genomes (KEGG) (Kanehisa and Goto, 2000) annotations. Bowtie 2 (Langmead and Salzberg, 2012) was used to map reads against predicted proteins from contigs to estimate abundance. Unannotated diversity was clustered with the cd-hit (Li and Godzik, 2006) at 70% identity and was used to assess metagenomic diversity. Metagenome sequences from each sample were mapped against bacterial reference genomes (O'Leary et al., 2016) using the NUCmer script from MUMmer (Marçais et al., 2018). Detailed statistical and bioinformatic methods are available at GitHub (https://github.com/genomica-fciencias-unam/squash-microbiome).

## 2.8 Data analyses

We used R for all diversity and statistical tests (R Development Core Team, 2011), and graphs were plotted with ggplot2 (Wickham, 2016). For biodiversity calculations, we used phyloseq (McMurdie and Holmes, 2013) and vegan (Oksanen et al., 2019). To evaluate the relationship between microbial communities with pH, Al, and nutrient content of soils, we calculated a constrained analysis of principal coordinates (CAP) (Anderson and Willis, 2003). Differential features of both taxa and predicted proteins were calculated using DESeq2 (Love et al., 2014). Bacterial genera were correlated with phenotypic traits using the psych R package (Revelle, 2017). Pathway reconstructions were created using the KEGG mapper (Kaneisa and Sato, 2020) with whole predicted proteins annotated in KEGG and those that were unique or overrepresented via DESeq2 analysis. Detailed statistical and bioinformatic methods are available at GitHub (https://github.com/genomica-fciencias-unam/squash-microbiome).

2.9 Data availability

Raw sequence data is available in the NCBI Sequence Read Archive (SRA) database. Soil sample sequence data is available with accession codes from SRR11972628 to SRR11972633; rhizosphere and endosphere sample data accessions from SRR11967772 to SRR11967813; and whole shotgun metagenomes accessions from SRR12046619 to SRR12046624.

# 3. Results and discussion

3.1 *Cucurbita root* microbiome is consistent with the two step-model for plant root microbiome acquisition

The 16S rRNA gene sequencing got us 7,476,012 paired sequences spanning $2.75 \times 10^9$ bp, merged sequences clustered into 73,277 OTUs (Table S2). Shotgun metagenomic sequencing yielded $247 \times 10^6$ sequences, assembled into $20 \times 10^6$ contigs representing $5.19 \times 10^9$ bp, an average GC% of 62.7, and N50 of 894 bp (Table S3).

We calculated the average 16S rRNA gene diversity indexes. In most cases, soil diversity values were greatest (Shannon = 7.31-7.99, Simpson = 0.996-0.999), followed by rhizosphere (Shannon = 3.9-7.31, Simpson = 0.948-0.996), and then root endosphere (Shannon = 3.9-7.2, Simpson = 0.88-995) (Fig. 1A, Table S2). Shannon and Simpson diversity indices showed statistically significant differences between soil and rhizosphere ($p < 1e-3$), soil and endosphere ($p < 1e-4$), but not between rhizosphere and endosphere. Thus, reducing diversity from soil to root observed in the squash microbiome is consistent with the two step-model (Bulgarelli et al., 2013).

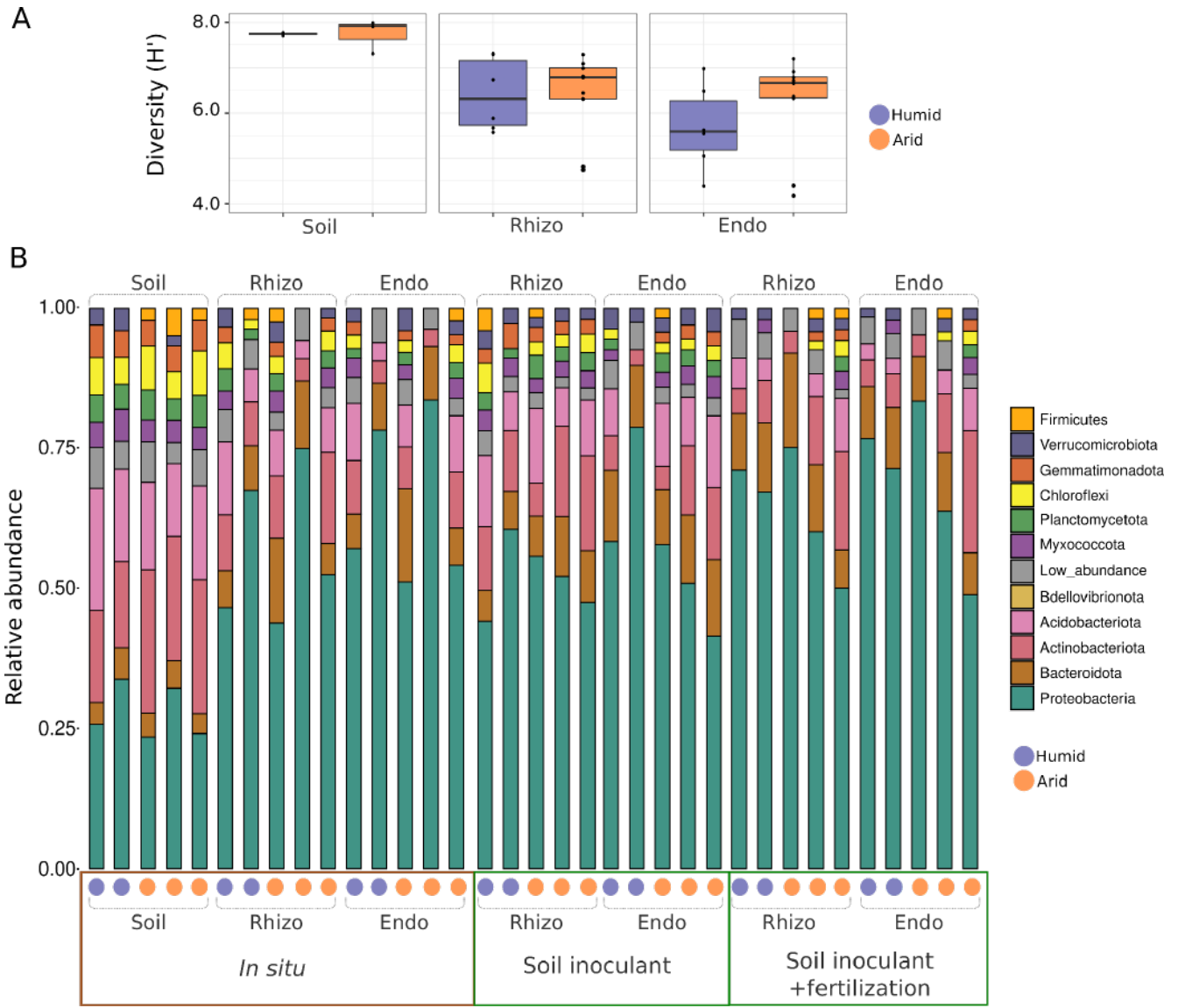

**Figure 1. Squash root microbiome diversity and phyla abundance in humid and arid conditions.** A) Shannon diversity index of the soil, rhizosphere (rhizo), and root endosphere (endo). B) Most abundant phyla are sorted by climatic conditions (humid or dry) of the soil, rhizosphere, and root endosphere from *in situ* collected plants and the common garden experiment (soil inoculants).

Besides different diversity levels, we found 74 genera exclusive from the rhizosphere and 70 genera exclusive from the root endosphere (Table S4). Endosphere exclusive genera include the intercellular plant pathogen *Pectobacterium* and the root isolated endophyte *Rothia* (Aizawa, 2014; Xiong et al., 2013). Proteobacteria was the most abundant phylum in all samples assessed (Fig. 1B), particularly abundant in root endospheres (average = 76%) and rhizospheres (69%), while its relative abundance was lower in soils (32%). In addition to *Cucurbita pepo*, plants like *Medicago sativa, Solanum lycopersicum, Lotus corniculatus, Oryza glaberrima*, and *Oryza sativa* have an increased relative abundance of Proteobacteria associated with their rhizospheres and root endospheres, which correlated with reduced levels of diversity (Bulgarelli et al., 2012; Fitzpatrick et al., 2018; Lundberg et al., 2012; Barajas et al., 2020).

3.2 Historically arid and humid soils contain different bacterial communities

Proteobacteria were increased in samples from historically humid origins for both *in situ* and the common garden experiment, while the Gram-positive Actinobacteriota and Firmicutes were elevated in historically arid conditions (Fig. 1B). We found significant differences (t-test, $p \leq 0.05$) in the proportion of Actinobacteriota (9% on average in arid samples and 5% in humid samples) and Firmicutes (1.3% in arid samples and 0.7% in humid samples) from the root endosphere. The rise of Gram-positive bacteria (Actinobacteriota and Firmicutes) in arid localities and roots resembles previously reported community compositions (Neilson et al., 2017). The greater abundance of Gram-positive bacteria in arid sites could be because they have a thicker cell wall that makes them resistant to desiccation, and some taxa are spore-forming microorganisms (Naylor et al., 2017).

A total of 1,009 bacterial genera were classified, with 199 exclusively arid conditions, 77 from humid climates, and 733 shared in both climates (Fig. S2). Some abundant genera from arid sources, such as *Yaniella, Woeseia, Salinimicrobium*, were described as halotolerant (Du et al., 2016; Li et al., 2004; Nedashkovskaya et al., 2010), suggesting the specialization of these communities. We calculated significant differential OTUs via DESeq2 analysis from historically humid and arid conditions ($p \leq 0.05$, Benjamini-Hochberg). We found genera distinctly identified from each condition like *Arenimonas, Azoarcus, Algoriphagus,* and *Thauera* (Tables S5 and S6). In arid climates, *Arenimonas* and *Azoarcus* were identified from root endospheres (Table S5). Previous reports have shown that *Azoarcus* can colonize roots (Chen et al., 2015), and some species are capable of nitrogen-fixing (Reinhold-Hurek et al., 1993). We expected the presence of *Arenimonas* since previous reports of isolates from soil and rhizosphere (Aslam et al., 2009; Kanjanasuntree et al., 2018). *Algoriphagus* and *Thauera* were enriched in rhizospheres (Table S6), but there are no previous reports in rhizospheres. However, some *Algoriphagus* were isolated from contaminated soils (Young et al., 2009), and *Thauera* can break down aromatic compounds (Breinig et al., 2000). Thus, both genera may play essential roles in promoting *Cucurbita* plant growth in polluted soils or breaking down aromatic root exudates.

### 3.3 The common garden experiment recovers *in situ* bacterial diversity

The Actinobacteriota increase in historically arid sites was found in *situ* and common garden experiments samples (Fig. 1B). A multidimensional scaling (MDS) using unweighted Unifrac distances (Lozupone et al., 2011) revealed discrete groups between historically arid and humid microbiomes. Splitting of arid and humid samples in the ordination was observed for both *in situ* and common garden samples, without differences between both groups (Fig. 2), supported by PERMANOVA analysis (999 permutations, $p < 0.001$). Although fertilizer treatments reduced microbial diversity (Table S2) as previously reported (Ling et al., 2017), changes in community structure promoted by fertilizer treatments were not abrupt since fertilized communities from arid and humid sources maintained their differences (Fig. 2). Finally, sterilized controls split apart from the rest of the samples, reinforcing the idea that the differences in bacterial communities originate from the source soil, and bacterial diversity in sterilized controls is mainly the result of greenhouse management (Fig. 2). These results suggest that small inoculates of soil microbial communities from arid sites can be exported to other systems to recover key taxa adapted to drought, with communities that retain their characteristics to an extent in response to *ex-situ* features as nutrient supply.

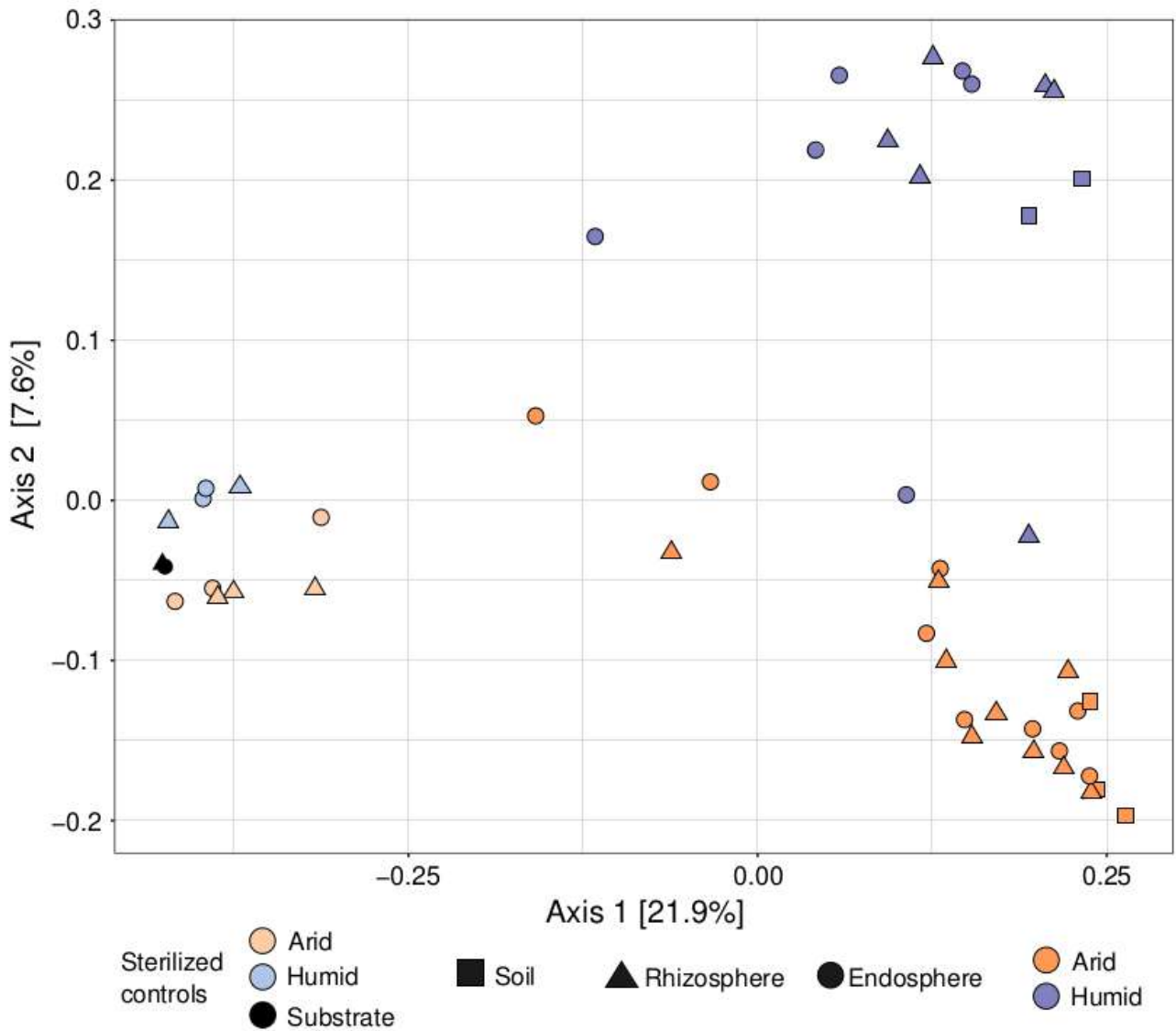

**Figure 2. Squash microbiomes split according to their source environment, but the common garden and *in situ* samples are mixed.** The -diversity is shown as unweighted Unifrac distances and plotted as multidimensional scaling (MDS) ordination, using the 16S rRNA gene amplicons data. Sterilized controls clustered at the left splitting from the soil inoculants and their soil and rhizosphere (21.9% variance, Axis1). MDS Axis 2 splits the arid and humid microbiomes (7.6%). The samples of soil, rhizosphere, and endosphere communities were closer to each other, regardless of *in situ* or common garden experiment samples.

Constrained analysis of principal coordinates (CAP) for in situ collected samples was built based on Al the physicochemical characterization of soils (Table 1). This analysis revealed that differences between humid and arid localities communities are related to Al and pH (Fig.

S3). The pH is a good predictor of bacterial community type because living in an extreme pH environment limits survival (Kaiser et al., 2016). Communities that live under pH stress tend to harbor phylogenetically similar organisms (Tripathi et al., 2018) such as Acidobacteriota, whose relative abundance increases with soil acidity (Lauber et al., 2009). A recent work that assessed global bacterial distribution patterns determined that pH, aridity, and ecosystem productivity are the best predictors of diversity (Delgado-Baquerizo et al., 2018).

Constrained analysis of principal coordinates (CAP) for in situ collected samples was built based on AI the physicochemical characterization of soils (Table 1). This analysis revealed that differences between humid and arid localities communities are related to AI and pH (Fig. S3). The pH is a good predictor of bacterial community type because living in an extreme pH environment limits survival (Kaiser et al., 2016). Communities that live under pH stress tend to harbor phylogenetically similar organisms (Tripathi et al., 2018) such as Acidobacteriota, whose relative abundance increases with soil acidity (Lauber et al., 2009). A recent work that assessed global bacterial distribution patterns determined that pH, aridity, and ecosystem productivity are the best predictors of diversity (Delgado-Baquerizo et al., 2018).

A constrained analysis of principal coordinates (CAP) ordination of the common garden experiment was controlled by plant phenotypic traits (Table S1) and supported the same pattern observed in MDS (Fig. S3). However, the CAP ordination allowed us to associate microbiome differences with plant phenotype, showing that microbiome samples from arid environments split into a discrete group with high aridity levels correlated to plant chlorophyll and carotenoid concentrations (Fig. S3).

## 3.4 Connecting root 16S rRNA gene microbiome diversity, shotgun metagenomics, and the *C. pepo* phenotype

Interestingly, some differentially (DESeq2) abundant genera comparisons between climates (Tables S5 and S6) correlated with plant phenotypic features (Table S1, Fig. 3). Statistically significant differences in Spearman's correlations ($p \leq 0.05$) revealed 25 genera of the rhizosphere and root endosphere correlated with plant phenotypes assessed (Fig. 3; Fig. S4).

The abundance of the genus *Cellvibrio* positively correlates with the foliar surface (r = 0.7), aerial length (r = 0.7), stem diameter (r = 0.7), total chlorophyll (r = 0.8), carotenoid levels (r = 0.7), and aerial biomass (r = 0.7) (Fig. 3; Fig. S4). *Cellvibrio* is common to plant inhabitants (Alcaraz et al., 2018; Jones et al., 2019; Zhang et al., 2020) with nitrogen fixation activity (Suarez et al., 2014) that may promote the increase of shoot biomass (Anderson and Habiger, 2012). Similarly, the relative abundance of *Ensifer* positively correlated with aerial biomass (r = 0.7), stem (r = 0.8), and aerial length (r=0.7) (Fig. S4). *Ensifer adhaerens* promote plant growth by secreting phytohormones and siderophores and solubilize phosphate (Oves et al., 2017). *Ensifer* was found in all root samples (rhizosphere and endosphere), highlighting this genus's relevance for *C. pepo*. *Acidovorax* was also ubiquitous and positively related to plant phenotype (r = 0.7 for carotenoids concentration) but enriched in humid samples (Fig. S4). Experimental evidence shows that *Acidovorax radicis* influences root and shoots weight of barley (Li et al., 2012). Thus, community members from arid locations may have plant growth promotion potential and may be relevant to protect crops from climate change-induced drought.

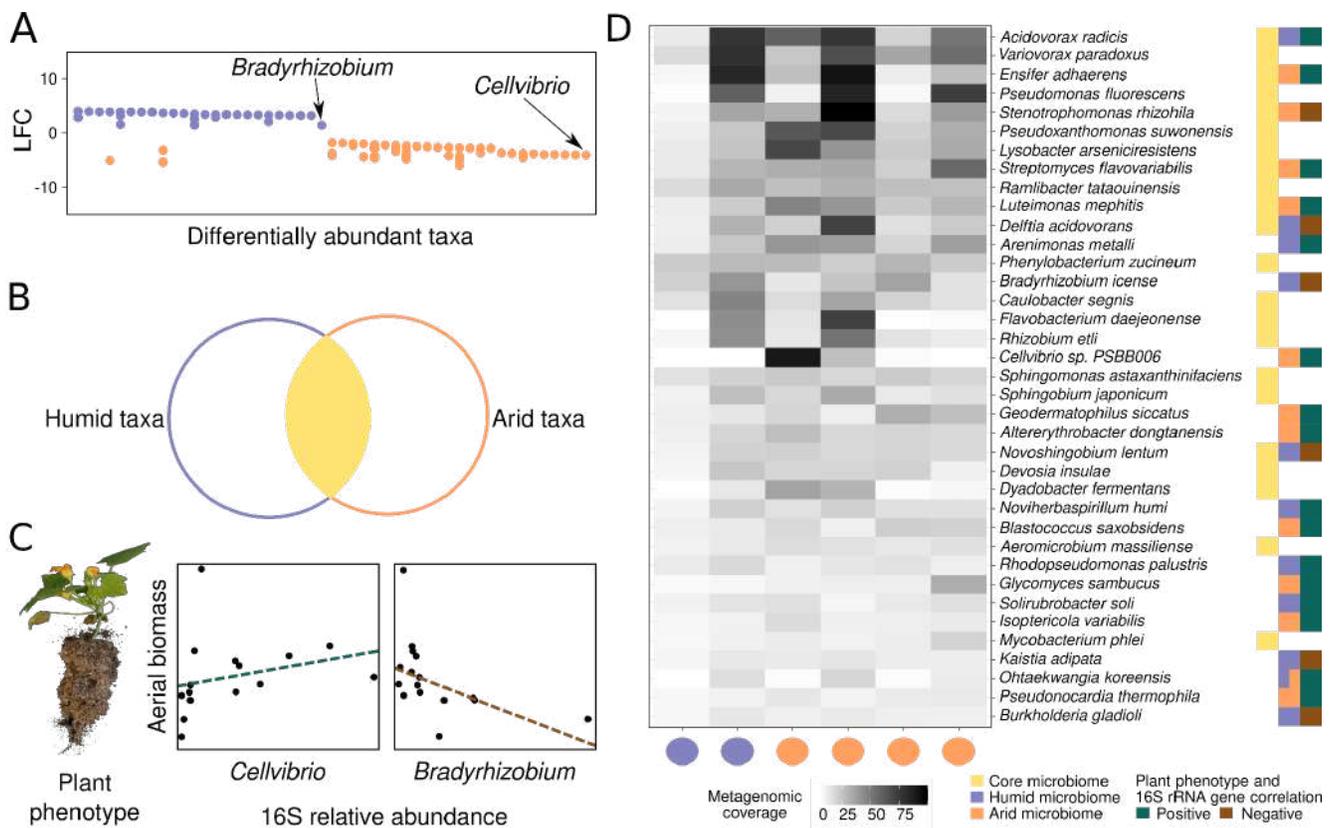

**Figure 3. Connections between plant phenotype, microbiome, and metagenomic recruitments.** A) Comparison and identification of significant differential OTUs (P ≤ 0.05, FDR) between arid or humid soils by contrasting their logarithmic fold change (LFC) ratio. The differential taxa are available in Tables S54 and S65. B) The squash core microbiome is defined as the intersection set of the common garden or *in situ* root-associated microbes (Table S76). C) Plant phenotypes were correlated with the relative abundance of genera enriched in arid or humid soils (Table S54 and S65) and plant phenotype (Table S2). D) Metagenomic binning of shotgun reads against reference bacterial genomes, in the heatmap, selected species of the same genera identified by 16S rRNA gene amplicons, but now identified up to species level, validating the 16S data. The metagenomic coverage is the percentage of the reference genome aligned against each metagenome. The color key indicates squash core microbiome affiliation, humid or arid 16S enrichment, and phenotype correlation in the left heatmap panel.

Surprisingly, *Methylobacterium* and *Novosphingobium* traditionally described as plant symbionts (Kutschera, 2007; Rangjaroen et al., 2017; Alcaraz et al., 2018) correlated negatively with some plant phenotypic traits (r = -0.7 and -0.6 for chlorophyll concentration and carotenoids levels in *Methylobacterium*; r = -0.8, -0.7, -0.7, and -0.7 for foliar surface, total chlorophyll, total carotenoids, and aerial biomass respectively in *Novosphingobium*) (Fig. S4).

Additionally, we looked in the shotgun metagenomes for sequences of further evidence and validation of the bacteria correlated with plant-phenotype features. Metagenomic recruitments against reference genomes produced evidence that supported findings of the 16S rRNA gene and allowed the identification up to species (Fig. 3). Some species belonging to genera detected in all the plant roots sampled microbiomes (Table S7), such as *Variovorax* and *Acidovorax,* were identified as *V. paradoxus* and *A. radicis*, were also identified in the shotgun metagenomic reads (Fig. 3). In arid environments, metagenomic sequences of binned bacteria confirmed microbiome significantly enriched ($p \leq 0.05$, Benjamini-Hochberg) OTUs like *Cellvibrio* sp. PSBB006, *Stenotrophomonas rhizophila*, *Streptomyces flavovariabilis*, *Arenimonas metalli,* and *Glycomyces sambucus* (Fig. 3). In humid environments by metagenomic binning, *Delftia acidovorans* and *Bradyrhizobium lablabi* were congruent with DESeq2 enriched OTUs (Fig. 3). The complete list of recruited genomes is available in Table S8.

Experiments using sorghum plants that analyzed the microbiome and transcriptome of root-associated bacteria revealed that plants secrete glyceraldehyde-3-phosphate to attract Actinobacteriota under drought conditions (Naylor and Coleman-Derr, 2018; Xu et al., 2018). Therefore, a pool of Actinobacteriota that are positively associated with plant traits may help plant survival under drought conditions. *Streptomyces, Glymocyces, Geodermatophilus,* and *Pseudonocardia* abundance are positively correlated with chlorophyll production. Further, a large number of metagenomic reads were binned as Actinobacteriota, remarkably *Streptomyces*. The high abundance of *Streptomyces* in our metagenomic sequencing data is outstanding but not surprising. Previous reports have described *Streptomyces* strains as plant growth-promoting bacteria that synthesize auxins, antibiotics, and siderophores (Sousa and Olivares, 2016), and their presence is associated with drought stress resistance (Fitzpatrick et al., 2018).

## 3.5 *Cucurbita pepo* microbiome genes from shotgun metagenomics reveal differences between humid and arid soil sources

We found coding genes and their predicted proteins used for annotation and further comparative analyses from the metagenomes. Using the Refseq derived annotations, we found a total of 1,192,784 predicted proteins from the squash metagenome, with 27,767 shared between all samples (Fig. S5). Then, using SEED classification (see Methods), we identified 2,969 predicted protein families detected in both climatic conditions, thus associated with squash in a climate-independent way. Then, 158 protein families were exclusively from humid sites, and 924 proteins were only found in arid localities (Table S9).

We compared significant differences ($p \leq 0.05$, Benjamini-Hochberg) in gene abundances between arid or humid metagenomes using DESeq2 analysis (Fig. S5; Tables S10 and S11). We found histidine kinase (BaeS), outer membrane protease (DegS), copper sensing regulator (CpxR), and sensory histidine kinase enriched in arid sites (QseC) (Table S10). These proteins are involved in sensing and responding to extracytoplasmic stress (De Wulf et al., 2002; Nagasawa et al., 1993; Walsh et al., 2003). The activation of responses to cell envelope damage may be essential to survive the loss of membrane integrity induced by desiccation (Lebre et al., 2017). The highly abundant miscellaneous category from arid sites included geranylgeranyl/isoprenyl reductase, phosphatidylglycerophosphatase, flotillin-like YqiK Lipid A export protein (Table S9 and S10). The proteins, as mentioned earlier, may be related to membrane function in extracytoplasmic stress (Bach and Bramkamp, 2015; Lin et al., 2016; Murakami et al., 2007). Concerning the balance of osmotic pressure, we determined that proteins from arid samples related to the synthesis of compatible solutes ectoine (EctA), hydroxyectoine (EctD), ABC transporters of osmolytes choline (OpuBB), and glycine-betaine (OpuAC), trehalose, and the N(+)/H(+) antiporter were significantly enriched or exclusive (Czech et al., 2018; Fig. S6; Tables S10, S11, and S12). The mycothiol synthase (MshD) and glutamate-cysteine ligase (involved in glutathione synthesis) were exclusive and enriched under arid conditions, respectively (Tables S9 and S10). Mycothiol and glutathione may be involved in oxidative stress resistance (Newton et al., 2008; Smirnova and Oktyabrsky, 2005) resulting from drought. Nutrient assimilation-related proteins enriched under arid conditions included xylanase and D-3 hydroxybutyrate oligomer hydrolase (Tables S9 and S10). These two enzymes are responsible for biopolymer breakage (Shirakura et al., 1983; Collins et al., 2005) and may degrade recalcitrant carbon sources abundant in dry oligotrophic soils (Naylor et al., 2017). The predicted proteins are part of molecular mechanisms associated with arid environments summarized into seven gene adaptations: stress sensing and response, protein degradation, osmoregulation, oxidative stress coping, cell wall, and catabolism (Fig. 4). Besides previously described categories, we predicted proteins related to chemotaxis, flagellar assembly, motility, and secretion systems type III, IV, and VI, which were identified independently of climatic conditions (Fig. S6). Chemotactic and flagellar proteins to sense and move toward plant signals and type III and IV secretion systems mediate interactions with host species (Hubber et al., 2004; Levy et al., 2018). Almost all enzymes involved in the nitrogen cycle were identified regardless of the climatic source of the samples (Fig. S6). In samples from arid sources, hydroxylamine levels could produce ammonia by the enzyme hydroxylamine reductase (Hpc), as suggested by the abundance of observed genes. In humid zones, hydroxylamine dehydrogenase (Hao) could dehydrogenate hydroxylamine to nitrite (Fig. S6).

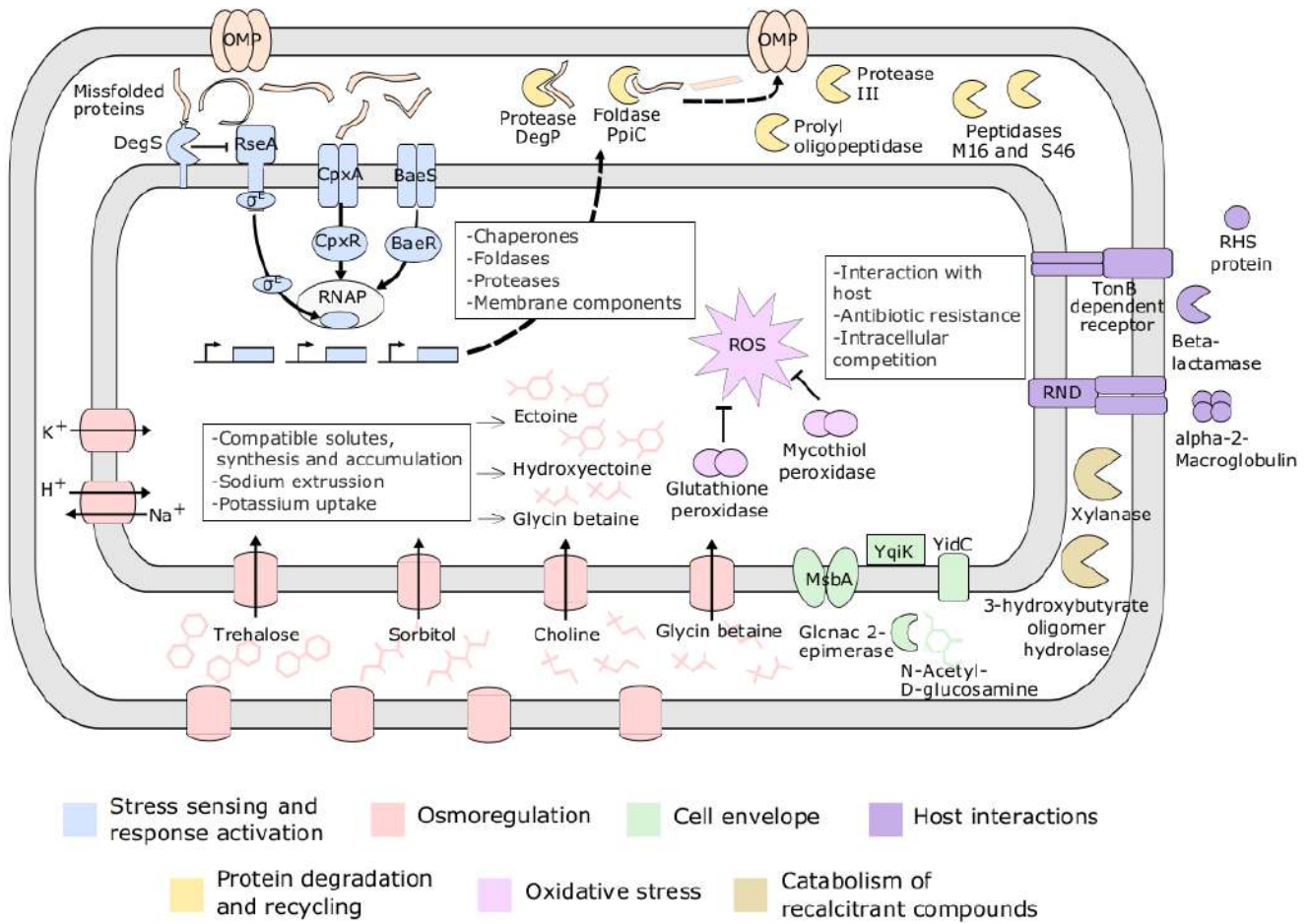

**Figure 4.** *Cucurbita pepo*'s **summary of microbiome adaptations to arid environments.** A metabolic proposal built from the main metagenomic predicted protein features. The main identified functional categories of the metagenomic adaptations are color-coded and classified based on their molecular or cellular functions.

Other enriched proteins in the arid environment include efflux transporter RND, beta-lactamase, membrane protein insertase (YidC), the RHS family, tex-like protein, alpha-2-macroglobulin family, and the TonB-receptor (Fig. 4 Fig. S5). Alpha-2-macroglobulin has been found in eukaryotic-interacting bacteria and probably inhibits host proteases (Budd et al., 2004). In contrast, Tex-like proteins are involved in toxin expression in some pathogens (Johnson et al., 2008). In addition to host interactions, RHS proteins mediate intracellular competition (Koskiniemi et al., 2013; Nikaido and Takatsuka, 2009), while beta-lactamases and RND efflux transporters promote antibiotic resistance. Other genes relevant to plant-bacteria interactions identified in the squash root metagenome were TonB receptors for iron siderophores and iron-chelating molecules that may drive bacteria-bacteria interactions (Kramer et al., 2020). Additionally, TonB-dependent receptors may be involved in plant carbohydrate assimilation (Blanvillain et al., 2007).

Further work will test the expression profiles of some genes mentioned here. We were cautious about metatranscriptomics approaches because of the less than 6.8 minutes as the half-life average for mRNAs operons under controlled conditions and single-species experiments (Selinger et al., 2003), having a complete profile with biological meaning would need sequencing depths out of the scope of this work. The significant differences between the arid and humid metagenomes are encouraging because of the selective pressure in bacterial genomes for size reduction (Mira et al., 2006; Pushker et al., 2004). Bacterial genome reduction is correlated with accelerated evolution (Dufresne et al., 2005), growth (Kurokawa et al., 2016), and some of the lost genes being beneficial in selected conditions but deleterious in alternative environments (Lee and Marx, 2012), suggesting that the differences in gene content could be adaptive for the tested conditions. We did not test genome reductions, but the same plants under contrasting conditions host differential taxa and genes, encouraging the microbiomes as diversity and genetic repositories for local adaptation.

## 4. Conclusions

Our experimental design allowed us to collect the *in situ* soil diversity, use it as a small inoculant for a common garden experiment while conserving their source traits. We identified the shared and differential root-associated bacteria taxa and genes in samples derived from arid and humid environments. Conserving the soil microbiota differences from humid and arid sites maintained in environmental samples and the common garden experiment was remarkable. This *ex-situ* microbiome conservation is especially notable since treatments involving soil inoculum consisted of a small fraction of substrates. Coupled with taxonomic differentiation, we detected coding genes associated with adaptations to arid environments. We described bacteria and genes relevant to plants' interaction and promoting their growth while providing a pool of metabolic abilities for coping with hydric stress.

# Declaration of competing interest

The authors declare no competing interests.

## Acknowledgments

This work was supported by the *Consejo Nacional de Ciencia y Tecnología* (Conacyt) *Problemas Nacionales* 247730 to DP. Conacyt *Ciencia Básica* 237387 and *Universidad Nacional Autónoma de México* DGAPA-PAPIIT UNAM IN221420 to LDA. Conacyt Ph.D. scholarship (CVU 742786) to CHA. Soil physicochemical analyses were carried out by Rodrigo Velázquez Durán. We appreciate field work assistance by Shamayim Martínez-Sánchez. Some analyses were carried out using CONABIO's computing cluster, which was partially funded by SEMARNAT through the grant *Contribución de la Biodiversidad para el Cambio Climático* with support from Alicia Mastretta-Yañez and Ernesto Campos. This paper is part of the doctoral thesis of CHA and is a requirement for obtaining a Ph.D. from the *Posgrado en Ciencias Biológicas, Universidad Nacional Autónoma de México* (UNAM). We thank the kind comments and suggestions of the reviewers that improved the manuscript.

## CRediT author statement

Using CRediT (Contributor Roles Taxonomy). Cristóbal Hernández-Álvarez: Conceptualization, Methodology, Visualization, Formal Analysis, Investigation, Writing-Original Draft. Felipe García Oliva: Conceptualization, Supervision. Rocío Cruz-Ortega: Investigation, Supervision, Resources. Miguel F. Romero: Formal analysis. Hugo R. Barajas: Resources, Formal Analysis. Daniel Piñero: Conceptualization, Funding acquisition, Resources. Luis D. Alcaraz: Conceptualization, Supervision, Funding acquisition, Project administration, Visualization, Writing - Reviewing and Editing.

## Supplementary figures

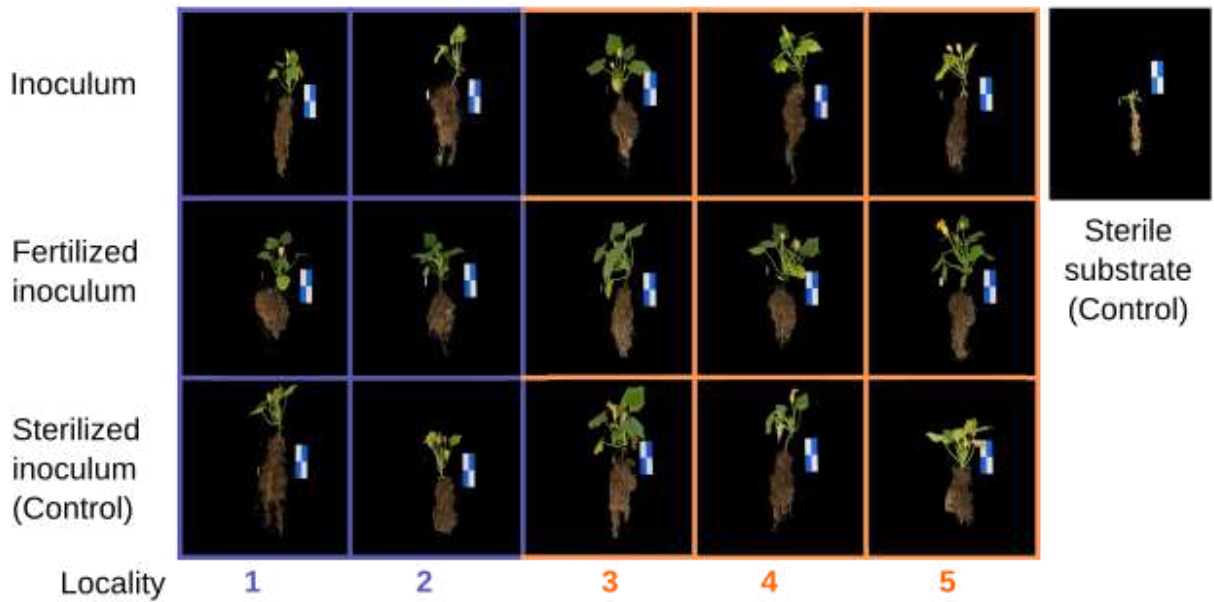

**Figure S1 Common garden experiment plants after 52 days.** The rectangles color indicates if the plants were grown on humid (blue) or arid (orange) inoculums. In the right upper corner, a plant grown without soil inoculum in sterilized substrate.

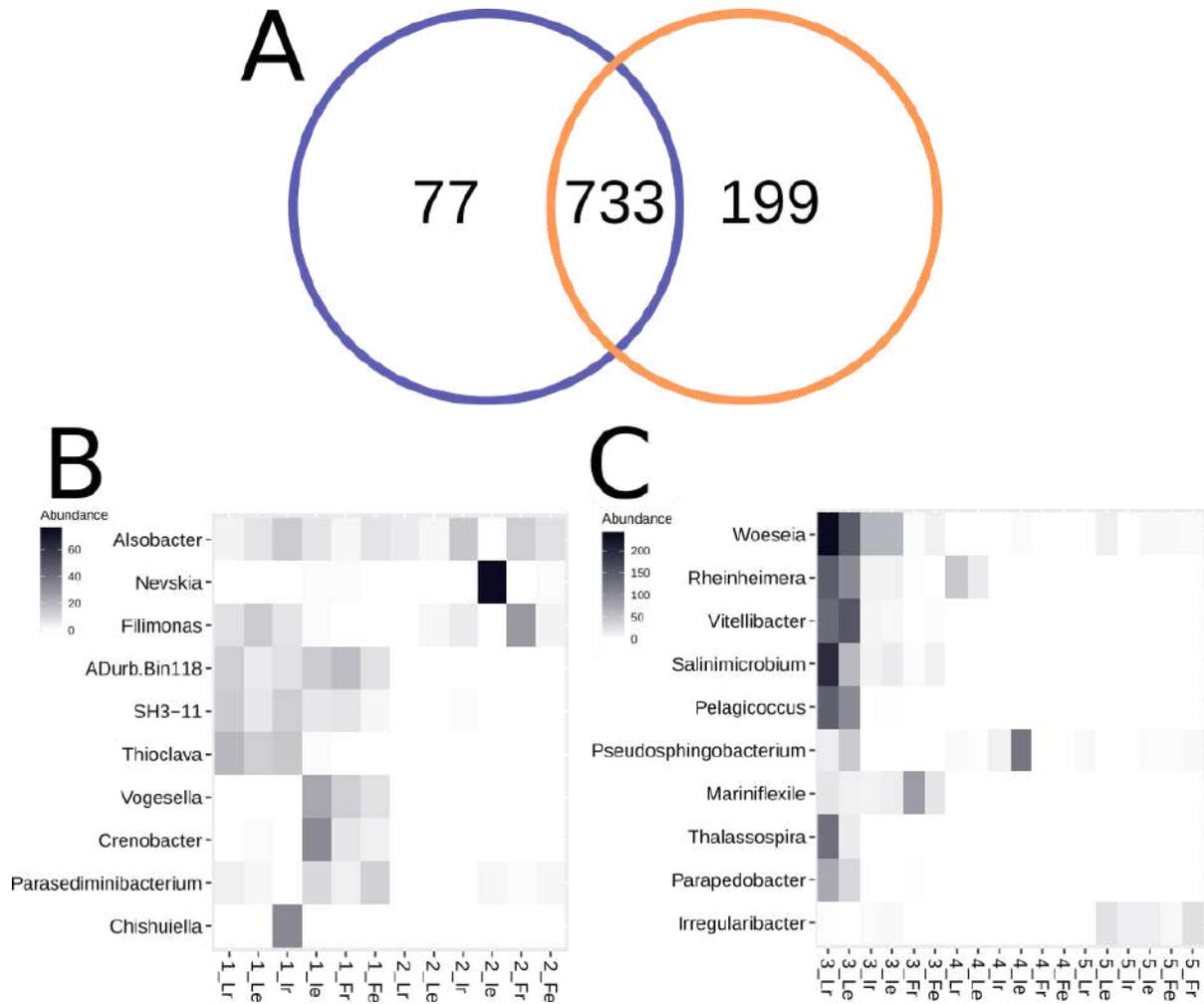

**Figure S2. Bacterial genera from historically arid and humid plant-associated samples.** A) Venn diagram contrasting both conditions. B) Most abundant genera from historically humid plant-associate samples. C) Most abundant genera from historically arid plant-associate samples.

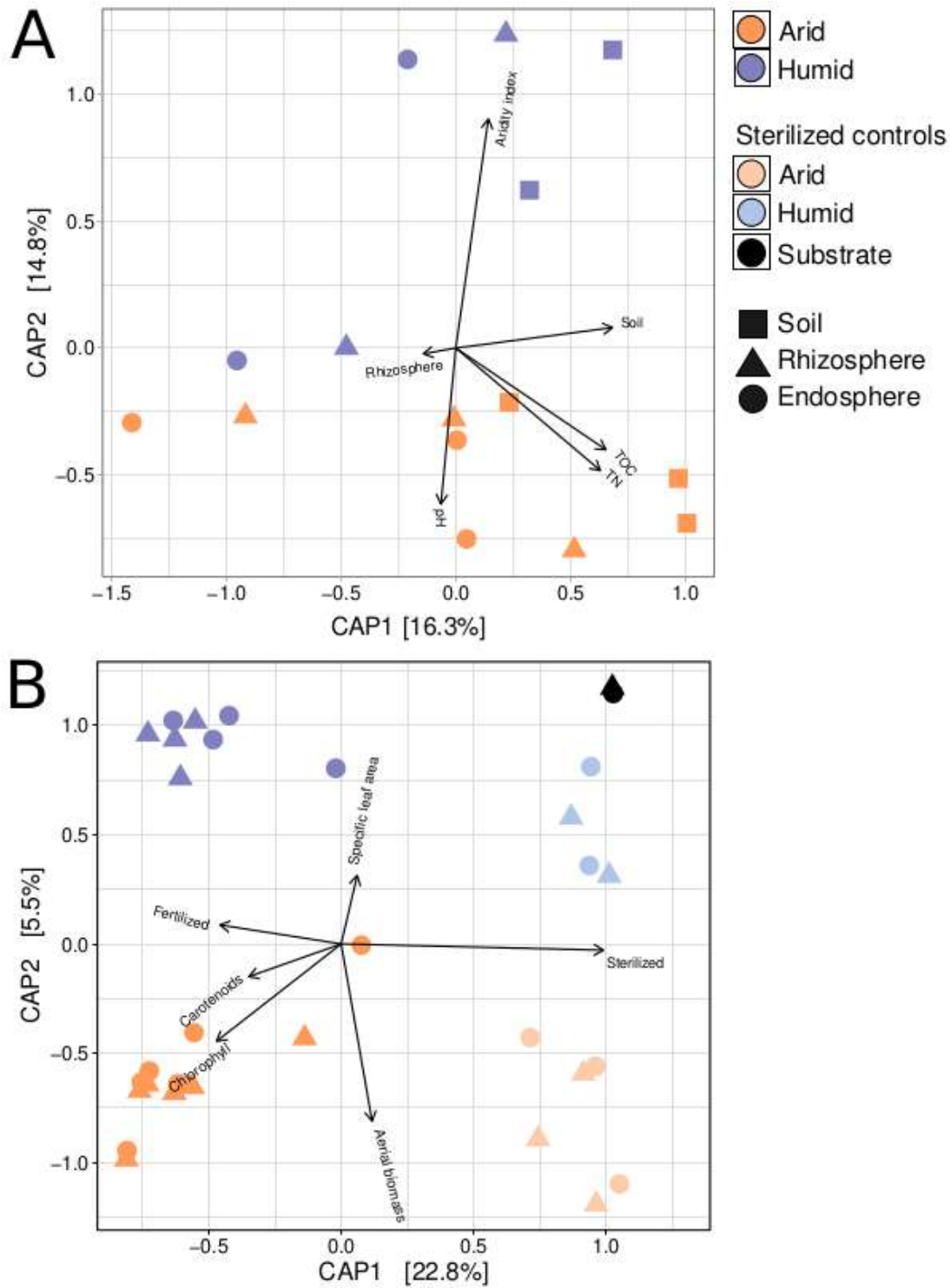

**Figure S3. Differences between bacterial communities originated from historically humid and arids localities.** A) CAP ordination based on unweighted Unifrac distances between common garden experiment samples and plant phenotype. B) CAP ordination of unweighted Unifrac distances from local source samples and environmental parameters (AI and physicochemical characterization of soils). Arrows indicate a correlation to variables. TN, total nitrogen; TOC, total organic carbon.

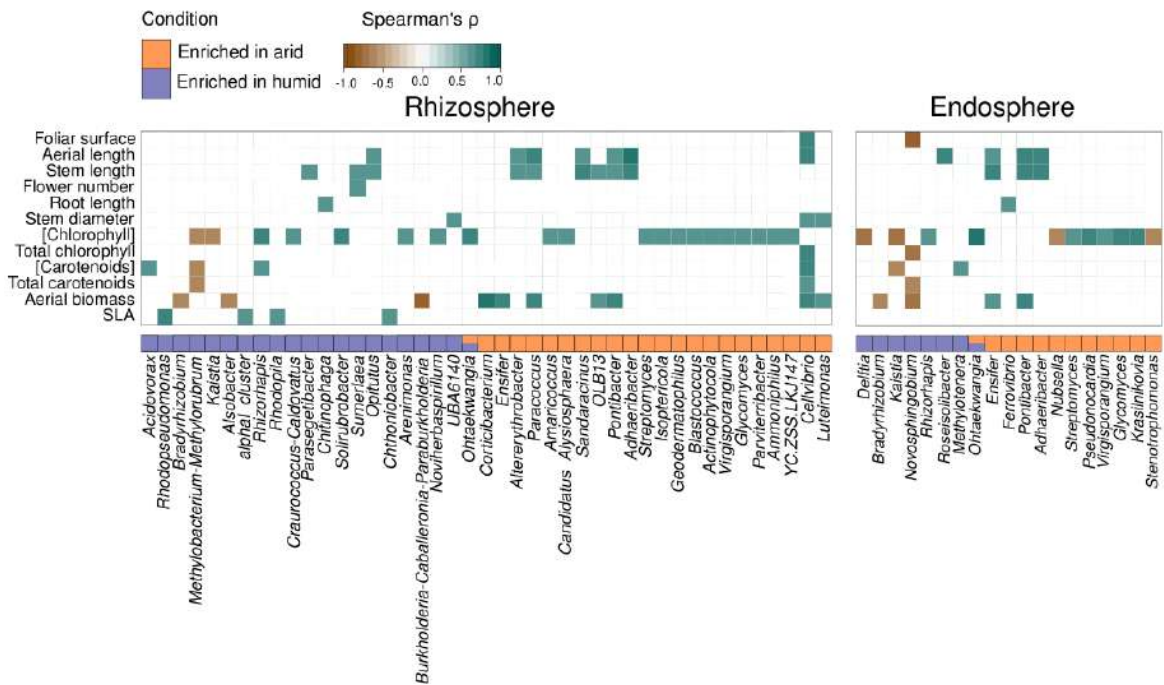

**Figure S4. Correlation between relative abundance of genera enriched in arid or humid soils and plant phenotype.** The correlations were evaluated for root-associated bacteria that were enriched in historically arid and humid soils (based on DESeq2 analyses). Only statistically significant values (P ≤ 0.05, FDR), not significant values are covered by a white square.

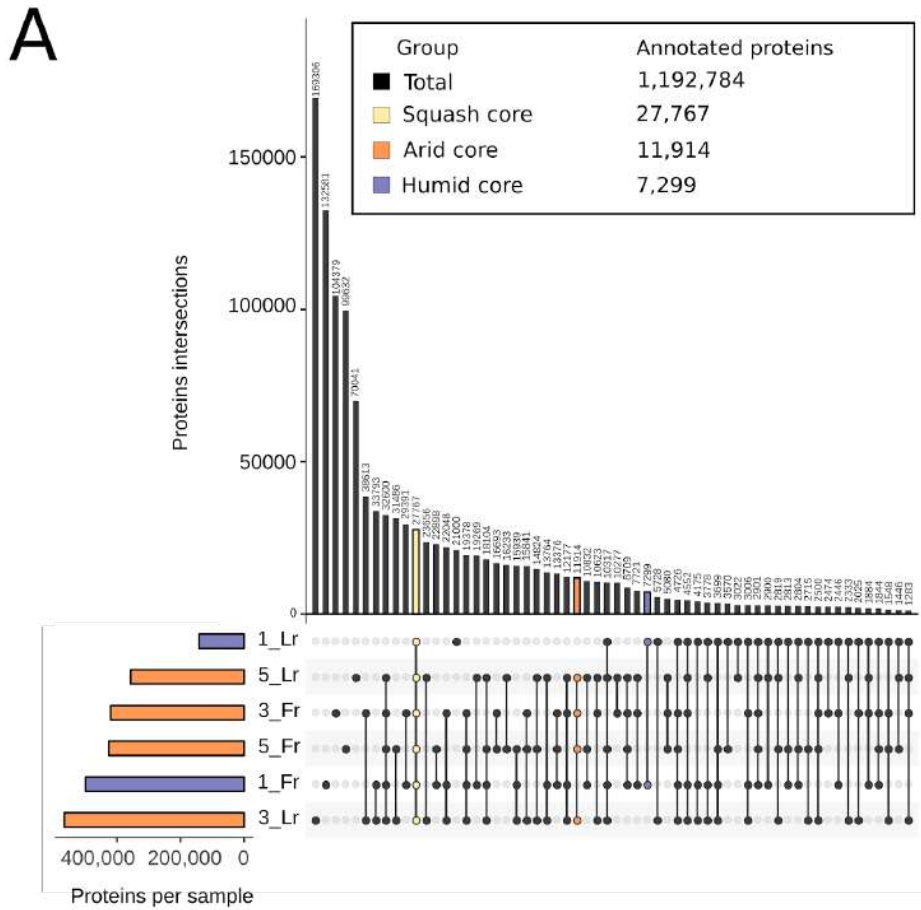
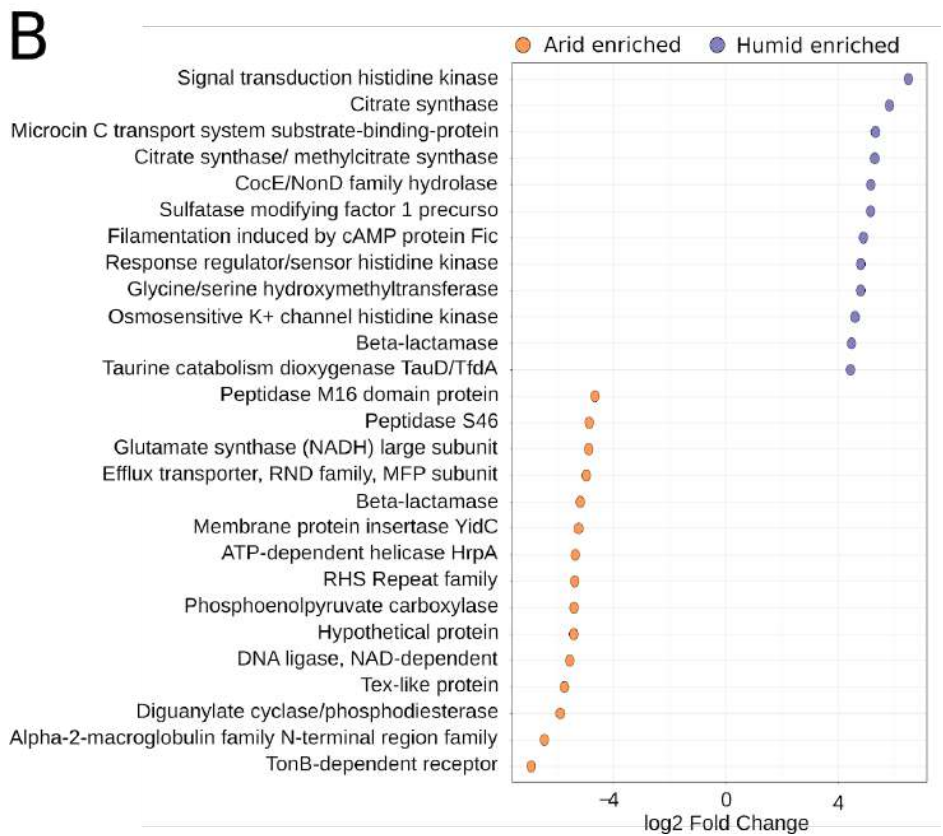

**Figure S5. Comparative analysis of predicted proteins (RefSeq) from root-associated communities recruited from historically humid and historically arid locations**. A) Upset plot contrasting predicted proteins according to the climatic source. Arid and humid cores are defined as the set of predicted proteins present in all samples from the same climatic source. Squash core refers to predicted proteins shared by the whole set of samples. B) DESeq2 analysis between both climatic conditions from squash core metagenome (p ≤ 0.05).

**A** Glycine, serine and threonine metabolism

**B** Phosphotransferase system (PTS)

**C** Nitrogen metabolism

**D** Bacterial chemotaxis

**E** Bacterial secretion system

**F** Flagellar assembly

**Figure S6. Metabolic pathway reconstructions of predicted proteins, using KEGG Mapper**. A) Glycine, serine and threonine metabolism. B) Phosphotransferase system. C) Nitrogen metabolism. D) Bacterial chemotaxis. E) Bacterial secretion system. F) Flagellar assembly. Core humid arid was defined as the set of proteins shared between both climatic conditions of source samples. Enriched proteins were determined according to DESeq2 analysis.

## Supplementary tables

Upon request.

**Table S1** Plant phenotypic traits evaluated in the common garden experiment. Phenotypic traits evaluated in the common garden experiment. Names are composed of *in situ* soil sampling localities (1-5) and treatment: soil inoculants (I), fertilized soil inoculants (F), sterilized soil controls (Z), and no soil control (C).

**Table S2** Sequencing, richness and diversity from 16S rRNA V3-V4 sequences. Names are composed by the number of localities, treatment and sample type. Localities are numbered from 1-5 from the most arid to the most humid. Treatments are local source samples (L), soil inoculum (I), fertilized inoculum (F), sterilized inoculum (Z) and controls without soil (C). Sample types are soil (s), rhizosphere (r), endosphere (e) and without plant (n).

**Table S3** Metagenomic sequencing, filtering, assembly and diversity of predicted proteins. Names are composed by the number of localities, treatment and sample type. Selected

localities were 1, 2, 3. Treatments are local source samples (L) and fertilized inoculum (F) from rhizospheres (r).

**Table S4** Comparative analysis between shared and unique genera from the rhizosphere and the root endosphere

**Table S5** Differential (DESeq2) endosphere OTUs in humid-arid comparisons.

**Table S6** Enriched (DESeq2) rhizosphere OTUs in humid-arid comparisons.

**Table S7** Cucurbita pepo core microbiome

**Table S8** Percentage of metagenomic sequences aligned against reference genomes. Reference genomes are sorted according to the average percentage of sequences aligned against mategenomes. Only reference genomes aligned with an average percentage of aligned sequences greater than 5% are shown.

**Table S9** Exclusive Subsystems in whole metagenomic sequences from rhizosphere according to climatic origin from soil samples.

**Table S10** Differential (DESeq2) humid-arid comparison rhizosphere genes (Subsystems).

**Table S11** Differential (DESeq2) humid-arid comparison rhizosphere genes (Kegg Orthologs).

**Table S12** Exclusive Kegg Orthologs (KOs) in whole metagenomic sequences from the rhizosphere according to climatic origin from soil samples.